\documentstyle[12pt,psfig]{ioplppt}

\begin{document}
\jl{1}
\bibliographystyle{plain}

\title{Boundary critical behaviour of two-dimensional random Ising models}[Random Ising models]
\author{F Igl\'oi\dag\ddag, ~ P Lajk\'o\ddag, ~ W Selke\S ~ and F Szalma\ddag}
\address{\dag\ Research Institute for Solid State Physics, H--1525 Budapest,
P. O. Box 49, Hungary}
\address{\ddag\ Institute for Theoretical Physics, Szeged University, H--6720 
Szeged, Hungary}
\address{\S\ Institut f\"ur Theoretische Physik, Technische
Hochschule, D--52056 Aachen, Germany}

\begin{abstract}
Using Monte Carlo techniques and a star--triangle transformation, Ising
models with random, 'strong' and 'weak', nearest--neighbour ferromagnetic 
couplings on a square lattice with a (1,1) surface are studied near
the phase transition. Both surface and bulk critical properties are 
investigated. In particular, the critical exponents of the surface
magnetization, $\beta_1$, of the correlation length, $\nu$, and of the
critical surface correlations, $\eta_{\parallel}$, are analysed.
\end{abstract}
\maketitle

\section{Introduction}
Quenched randomness may have a profound effect on the
nature of phase transitions. If there is a continuous phase transition
in the perfect system then, according to the Harris criterion\cite{harris},
the relevance of the perturbation is connected to the sign of
the specific heat exponent $\alpha$ in the pure system. The two-dimensional
random Ising model with $\alpha=0$ represents the
borderline case of the
perturbational theory. Indeed, that model has been
the subject of intense investigations to clarify its critical properties
\cite{dot1,sha1,sel1}. 

According to field-theoretical studies\cite{dot1,sha1} the
randomness is, in the renormalization
group sense, a marginally irrelevant perturbation, therefore it leads to
logarithmic corrections to the power-law singularities of the pure model.
For example, the bulk magnetization, $m_b$, and the correlation length,
$\xi$, are expected to behave near the transition point as
\begin{equation}
m_b\sim t^{1/8} |\ln t|^{-1/16}\;,
\label{mb}
\end{equation}
and
\begin{equation}
\xi\sim t^{-1} |\ln t|^{1/2}\;,
\label{xi}
\end{equation}
where $t=|T_c-T|/T_c$ is the reduced temperature.
The critical spin-spin correlation function $G(r)$ {\it averaged} over several
samples has a pure power law decay \cite{dpp}
\begin{equation}
G(r) \sim r^{-1/4} \left[ {\cal A} + {\cal B} / (\ln r)^{-2} \right]\;,
\label{avrcorr}
\end{equation}
whereas the {\it typical} correlation function calculated in a large single
sample is conjectured\cite{ludwig} to decay as
\begin{equation}
G(r) \sim r^{-1/4}\left( \ln r \right)^{-1/8}\;.
\label{Gr}
\end{equation}
The above conjectures are found to be in agreement with numerical results
of large--scale Monte Carlo (MC) simulations\cite{wang1,sel1,parisi} and
transfer matrix calculations\cite{quei1,stauffer,quei2}. However, also
conflicting interpretations of the numerical findings have been
suggested, invoking dilution--dependent critical exponents
and weak universality \cite{kim,kuhn}.

In this paper, we consider the boundary critical behaviour of the
two-dimensional random bond Ising model. The surface critical properties
of the perfect model are exactly known since many years \cite{mcwu}.
For example, the asymptotic behaviour of the surface magnetization, $m_1$,
and the correlation length, $\xi_{\parallel}$, measured
parallel to the surface, is given by
\begin{equation}
m_1 \sim t^{1/2}
\label{m1}
\end{equation}
and
\begin{equation}
\xi_{\parallel} \sim t^{-1}\;,
\label{xipar}
\end{equation}
whereas the critical surface spin-spin correlation function has the
asymptotic decay:
\begin{equation}
G_s(r) \sim r^{-1}\;.
\label{Gspure}
\end{equation}
Thus the corresponding critical exponents are $\beta_1=1/2$,
$\nu_{\parallel}=1$
and $\eta_{\parallel}=1$. No field-theoretical results are available for
the random case. However, it seems reasonable to expect, in analogy to the
bulk properties, that the randomness is a marginally irrelevant variable
at the surface fixed point as well. Then one might obtain
logarithmic corrections to the asymptotic behaviour of
the perfect model.

In the present study, we performed extensive numerical
investigations to illuminate this
issue by determining the surface critical properties of the
Ising model with nearest-neighbour random couplings on the square lattice.
In our first approach, we used large-scale MC techniques and computed the
surface magnetization and the complete magnetization profile of the model.
Our second method is based on the star-triangle (ST)
transformation. By that method we calculated both the surface magnetization
and the surface correlation function of the model. By the two, in several
respects complementary approaches, we determined numerically the complete
set of surface critical exponents, including the surface magnetization
exponent $\beta_1$, the correlation length exponent $\nu_{\parallel}$ and
the decay exponent of the critical suface correlations $\eta_{\parallel}$.
Note that some of the MC results on the surface magnetization
have already been announced in a short communication\cite{ssli}.

The paper is organised as follows. The MC results on
the surface magnetization and the magnetization profiles are presented
in Section 2. In Section 3 we describe the ST approach as applied to
the random Ising model and discuss the numerical results on the
surface magnetization
and the surface correlation function. The main conclusions are given in the
Summary. Some details of the ST method have been transferred to the
Appendix.

\section{Monte Carlo simulations}
Let us consider the Ising model with nearest-neighbour 
ferromagnetic couplings, where
the spins $s_{i,j}$(=$\pm 1$) are situated
on the sites $(i,j)$ of a square lattice. A surface may be
introduced by cutting the coupling bonds 
along one of the axes of the lattice, leading to the
(10) surface, or along the diagonal, leading to the
(11) surface. In the MC simulations, we studied systems
with two parallel surface lines, each line having $L$ sites. Each row
perpendicular to the surface consists of $K$ sites. The spins in the
first and last row are assumed to be connected by periodic boundary 
conditions. The lines parallel to the surfaces are numbered by the
index $i$, i.e. $i = 1$ and $i = K$ denote the two surface lines.
The index $j$ refers to the position along a line, running from 
1 to $L$. The total number of spins is $K \times L$. The
aim of the simulations is to determine thermal properties
of the semi--infinite system, where $K, L \longrightarrow \infty$;
therefore finite--size effects need to be studied with care.

The interaction between neighbouring spins may be either
'strong', $J_1 > 0$, or 'weak', $0 < J_2 < J_1$. Strong and weak
couplings are  distributed randomly, with $p$ (or $1- p$) being the
probability of a weak (or strong) bond. If both interactions occur
with the same probability, $p = 1/2$, then the model is
self--dual \cite{fisch}. The self--dual point is located at

\begin{equation}
\tanh(J_1/k_BT_c) = \exp(-2J_2/k_BT_c).
\label{self-dual}
\end{equation}
determing the critical temperature, $T_c$, of the
bulk Ising system ($K, L \longrightarrow \infty$, and
full periodic boundary conditions), if the
model undergoes one phase transition. Indeed, results of
previous simulations support that assumption \cite{sel1,wang1}. The
simulations were done for the self--dual case, i.e. at $p$= 1/2.

Certainly, one
expects that both bulk and surface will still order at the bulk critical
temperature, $T_c$, in a two--dimensional
model with short--range interactions: The one--dimensional
surface does not support any separate ordering, so that one encounters
the 'ordinary transition' \cite{bind1,diehl}. 

Varying, in the self--dual case,
the ratio of the strong and weak couplings, $r = J_2/ J_1$, one
may change the degree of dilution. At $r=1$, one recovers
the perfect Ising model, while $r = 0$ corresponds to the
percolation limit, where $T_c = 0$. As had been shown before, the
crossover to the randomness dominated bulk critical behaviour may
be monitored conveniently by choosing $r$ in the range
of 1/10 to 1/4. Then the crossover length, at criticality, ranges
from a few to about 20 lattice spacings \cite{wang1}. Indeed, we simulated
the random model at these two values, $r =1/4$ and 1/10, augmented 
by computations for the perfect model, $r = 1$.

Most of the simulations were performed for the (11) surface, albeit
a few runs were also done for the (10) surface to compare
with exact results. For the (11) surfaces, we  
usually set $L = K/2$, with
$K$ ranging from 40 to 1280 to check for finite--size effects.
For the (10) surfaces, quadratic systems were studied.
We averaged over an ensemble of bond configurations (or 
realizations). The number of realizations typically ranged from at least 15
for the largest systems up to several hundreds for
the small systems. In general,
the one--cluster flip Monte Carlo algorithm was used (mainly for
testing purposes, we also applied the single--spin flip
method), generating, close
to the critical point, several $10^4$ clusters per realization. Note
that the statistical errors for each realization were significantly
smaller than those resulting from the ensemble averaging. To avoid
inaccuracies due a, possibly, unfortunate choice of the random
number generators, we compared
results obtained from shift register and linear congruential
generators.

The crucial quantity, computed in the MC simulations, is the
magnetization profile. It is described by the
magnetization per line, $m(i) = < |\sum s_{i,j}| >/L$,
where $s_{i,j}$ denotes the spin in line $i$ and row $j$,
with $i = 1, 2,... K$, and summing over 
$j = 1, 2, ... L$. The absolute values are taken to
obtain a non--vanishing profile for finite systems, as usual.  
The surface magnetization is given by $m_1 = m(1)= m(K)$. 

Because the distribution of the random bonds is the same in the bulk and
at the surface, one may expect a monotonic decrease of $m(i)$ on 
approach to the surface, due to the reduced coordination number at 
the surface (being two for the (11) and three for the (10) surface).
This behaviour is illustrated in Fig. 1, comparing magnetization
profiles of the perfect, $r=1$, and random, $r =1/4$, Ising model with 
a (11) surface, at the same distances from $T_c$, measured by the
reduced temperature $t = |T - T_c|/T_c$. The critical point, $T_c$,
follows from (1). Obviously, randomness tends to suppress the
magnetization, at fixed value of $t$. The profiles display a pronounced
plateau around the center of the systems, at which the bulk
magnetization, $m_b$, is reached. The existence of the broad plateau
indicates that the linear dimension $K$ of the MC system is
sufficiently large to compute, for instance, the surface
magnetization of the semi--infinite system. Of course, in addition
one has to monitor possible changes of $m_1$ with $L$, to have 
possible finite--size effects due to that dimension under control.-- Note
that $m_b$ is known exactly in the
perfect case \cite{mcwu}, and very accurately in the random
case \cite{wang1}. 

For the perfect two--dimensional Ising model with a (10) surface, the
complete magnetization profile has been calculated
exactly
%F1
in the continuum limit
%F1 
\cite{bariev,czer}. In particular, the profile approaches
the bulk value in an exponential form, with $m_b - m(i) \propto
exp( - i/ \xi_r)$, where $\xi_r$ is the bulk 'correlation range',
which becomes only asymptotically, as $T \longrightarrow T_c$, 
identical to the bulk 'true correlation length' \cite{fisher}. Indeed,
we tested the accuracy of our simulational data by comparing them,
for the (10) surface, to the exact expression. In addition, we found
that the same correlation range determines the exponential approach 
of the magnetization towards its bulk value in the (11) case as well.

For the perfect two--dimensional Ising model with a (11) surface, exact
results exist for the surface magnetization, $m_1$, and the magnetization
in the next line, $m(2)$ \cite{peschel}. Again, the Monte Carlo
data, obtained with modest computational efforts,
agreed very well with the exact results, as shown in Fig. 2. In the figure,
the 'effective exponent' $\beta(i)_{eff}$ is depicted, defined by

\begin{equation}
\beta(i)_{eff}(t) = d \ln (m(i))/d \ln (t)
\label{betaeff}
\end{equation}

Certainly, as $t \longrightarrow 0$, the effective exponent acquires
the true asymptotic value of the critical exponent $\beta(i)$. For example,the
asymptotic critical exponent of the surface magnetization is $\beta(1) =
\beta_1 =1/2$, being, by the way, identical for (11) and (10)
surfaces. Because the magnetization $m(i)$ is computed at discrete
temperatures $t_k$, we use in analysing the simulational data, instead
of (9) the corresponding difference expression, with $t = (t_k + t_{k+1})/2$.
The error bars, included in Fig. 2, have been calculated in a conservative
fashion, getting the bounds for $\beta$ by comparing the upper (lower)
limit of $m(i)$ to the lower (upper) limit of $m(i+1)$, where the
bounds of the magnetization are computed in the standard way from the
ensemble averaging. Alternately, we also computed the error bars from
usual error propagation, which turned out to be appreciably smaller.  

In Fig. 2, the temperature dependence of the effective exponent 
$\beta(i)_{eff}$ deeper in the bulk
is also displayed. For example at $i= 10$, one readily observes 
the crossover from the bulk effective exponent 
(as follows from the exact expression for the bulk
magnetization \cite{mcwu}) to the surface
dominated behaviour, when the correlation length becomes large
compared to the distance from the surface. In general, at finite and
arbitrarily large distances to the surface, $\beta(i)_{eff}$ will
always converge, on approach to $T_c$, to 
the surface critical exponent, $\beta_1$ = 1/2,
and not to the bulk critical exponent, $\beta$ =1/8.
Analogous observations have 
been reported for three--dimensional Ising models with
surfaces \cite{pleim}.

The main aim of the Monte Carlo study has been to estimate $\beta_1$ in
the random case. Results of the extensive simulations are summarized in 
Fig. 3, depicting the effective exponent $\beta(1)_{eff}(t)$
at $r =1/4$ and $r=1/10$, compared to its exactly known form for the
perfect case, $r=1$. Typical error bars, increasing closer to criticality, are
displayed, based on standard error propagation resulting from the
variance in ensemble averaging of $m_1(t_k)$ and $m_1(t_{k+1})$. Data obviously
affected by finite--size effects have not been included in the figure.

As seen from Fig. 3, at fixed distance from the critical
point, $t$, $\beta(1)_{eff}$ rises systematically with increasing
dilution, reflecting the decrease in $m_1$ with stronger randomness.
However, asymptotically, $t \longrightarrow 0$, it is well conceivable that
the surface critical exponent will coincide in the perfect and dilute
cases, with $\beta_1 =1/2$. Indeed, a reasonable estimate, both for
$r=1/4$ and $r=1/10$, is $\beta_1$ =0.49 $\pm 0.02$. 

Thence, the simulations demonstrated that the critical exponent $\beta_1$ is
rather robust against introducing randomness
simultaneously in the bulk and at the surface. Note that $\beta_1$
remains 1/2 too, when only the surface bonds of the two--dimensional 
Ising model are randomized as described above, but
keeping a unique bulk coupling, as we confirmed in simulations.
Interestingly enough, in the three--dimensional case, introducing
random nearest--neighbour strong and weak surface bonds, but having
only one interaction for the bulk couplings, seems to be an irrelevant
perturbation as well, i.e. the surface critical exponent seems
to be the same as for the perfect
surface, $\beta_1 \approx 0.80$ \cite{pleim}. This
robustness may indicate that the bulk critical fluctuations play
a crucial role for the surface critical exponent, albeit it is
not determined by bulk critical exponents \cite{bind1,diehl}.
If that is true, then our result for the two--dimensional
case with random bulk and surface interactions
suggests that the bulk critical fluctuations
are not very sensitive towards dilution (in accordance with the
theory of, at most, logarithmic modifications of the asymptotic power--laws
describing critical behaviour of the perfect system
in two--dimensional Ising models \cite{sha1,sel1}). We shall come back to
this aspect in the next Section.

\noindent
\section{Star-triangle transformation}

The star--triangle transformation was introduced by Hilhorst and
van Leeuwen\cite{hvl}, and
used later by others\cite{bghvl,il,bg} to calculate the surface magnetization
and the surface correlations in layered triangular lattice Ising models.
Here we generalize the method for non-translationally invariant systems.

\subsection{Star-triangle approach to boundary behaviour}

The method is based on an exact mapping of the original triangular
model, with couplings $\{K_1\},\{K_2\}$ and $\{K_3\}$, to a
hexagonal model with couplings $\{p_1\},\{p_2\}$ and $\{p_3\}$ denoted by
dashed lines in Fig. 4. In the transformation
the right-pointing triangles are replaced by stars
such that the couplings are related by
\begin{equation}
K_1={1 \over 4} \ln \left({\cosh(p_1+p_2+p_3)\cosh(-p_1+p_2+p_3)
\over \cosh(p_1+p_2-p_3)\cosh(p_1-p_2+p_3)} \right)\;,
\label{transf}
\end{equation}
and its cyclic permutation in the indices $i=1,2,3$. In the second step
of the mapping the left-pointing stars of the hexagonal lattice are replaced
by triangles resulting in a new triangular lattice, which is denoted by
dotted lines in Fig. 4. Iterating this procedure a sequence of triangular
Ising models is generated ($n=0,1,2,\dots$) from the original model with
$n=0$.

As seen in Fig. 4, the surface spins of the $n$-th and the $(n+1)$-th
models are connected by the surface couplings of the intermediate hexagonal
model. In this geometry, the thermal average of the
$l$-th surface spin of the $n$-th model, denoted by $\langle s_l^{(n)} \rangle
\equiv \sigma_l^{(n)}$
is connected to the thermal averages of the neighbouring spins
$s_{l-}^{(n+1)}$ and $s_{l+}^{(n+1)}$ of the
$(n+1)$-th model, where the corresponding surface couplings of the hexagonal
lattice are denoted by $p_{l-}^{(n+1)}$ and $p_{l+}^{(n+1)}$.
As shown in the Appendix, one has
\begin{equation}
\sigma_l^{(n)} = a_{l+}^{(n+1)} \sigma_{l+}^{(n+1)} +
a_{l-}^{(n+1)} \sigma_{l-}^{(n+1)}\;,
\label{magnrel}
\end{equation}
where
\begin{equation}
a_{l+}^{(n)}=\tanh(p_{l+}^{(n)}) {1-\tanh^2(p_{l-}^{(n)})
\over 1-\tanh^2(p_{l+}^{(n)}) \tanh^2(p_{l-}^{(n)})}\;,
\label{a+l}
\end{equation}
while in $a_{l-}^{(n)}$ one should interchange $p_{l+}^{(n)}$ and
$p_{l-}^{(n)}$. Now using the
vector notation ${\bf \sigma}^{(n)}$ for the surface spin exceptational
values in the $n$-th model
and similarly ${\bf \sigma}^{(n+2)}$ for the $(n+2)$-th model we obtain the
relation
\begin{equation}
{\bf \sigma}^{(n)}= {\bf A}^{(n+1)} {\bf A}^{(n+2)} \bf{\sigma}^{(n+2)}
\label{vectrel}
\end{equation}
where the non-vanishing elements of the ${\bf A}^{(n+1)}$ matrix are given
by $a_+^{(n+1)}(l)$ and $a_-^{(n+1)}(l)$ in terms of the surface couplings
of the $(n+1)$-th
hexagonal lattice, equation (\ref{a+l}), and similar relation holds
for ${\bf A}^{(n+2)}$.
(We consider two successive steps in (\ref{vectrel}) in order to avoid
complications with the different parity of the odd and even number of
transformations.)
Now taking
the boundary condition $\lim_{n \to \infty} {\bf \sigma}^{(n)}=(1,,1,\dots,1)$
we obtain for the average surface magnetization
\begin{equation}
m_1=\lim_{L \to \infty} {1 \over L} \sum_{l=1}^L \sigma_{l}^{(0)}=
\lim_{n \to \infty} f(n)\;,
\label{ms}
\end{equation}
with
\begin{equation}
f(n)=\lim_{L \to \infty} {1 \over L} \sum_{i,j=1}^L \left[ \prod_{k=1}^n
\bf{A}^{(k)} \right]_{ij}\;.
\label{fn}
\end{equation}
We note that $f(n)$ in (\ref{fn}) is
formally equivalent to the partition function of an $n$-step directed walk
(polymer)
in a random environment, where the (random) fugacities corresponding to the
$k$-th step of the walk are contained in the ${\bf A}^{(k)}$ matrix, which is
just the transfer matrix of the directed walk.

Next we consider the average connected surface correlation function defined as
\begin{equation}
G_s(l)=\lim_{L \to \infty} {1 \over L} \sum_{i=1}^L \left[ \langle s_{i+l}
s_i \rangle - \langle s_{i+l} \rangle \langle s_i \rangle
\right]\;.
\label{corr}
\end{equation}
As shown in the Appendix, the surface correlations in the $n$-th triangular
model are connected to those in the $(n+1)$-th model, and
the relation is given in terms of the surface couplings of the intermediate
hexagonal lattice, equation (\ref{g1}), similarly
to (\ref{magnrel}). Furthermore,
as we argue
in the Appendix, in the asymptotic limit ($l \gg 1$) the surface correlation
function can be expressed by the partition function $f(n)$ of the corresponding
directed walk,
\begin{equation}
G_s(l) \propto \int_0^{\infty} d n {l \over n^{3/2}} \exp \left(-{l^2
\over n} \right)[f^2(n)-f^2(\infty)]\;.
\label{corr1}
\end{equation}
Thus the surface properties of the model are connected to the asymptotic
behaviour of $f(n)$ in (\ref{fn}). For different
temperatures, corresponding to different
thermodynamical phases of the random Ising model,
the $f(n)$ function has different asymptotic behaviour, as can be seen in
Fig. 5 for a dilution of $r=1/10$.

In the ordered phase, $T < T_c$, $f(n)$ approaches a
finite limit, the surface
magnetization $m_1$, through an exponential decay,
\begin{equation}
f(n)= m_1(T) + {\cal A} \exp(-n/\xi_{\parallel}^2)~~~~T < T_c\:.
\label{fn-}
\end{equation}
For $T \ge T_c$ the limiting value of $f(n)$ is zero, which corresponds to
vanishing surface magnetization, and the decay for $T>T_c$ is exponential,
\begin{equation}
f(n) \propto \exp(-n/\xi_{\parallel}^2)~~~~T > T_c\:,
\label{fn+}
\end{equation}
whereas {\it at the critical point}, it has the form of a power law
\begin{equation}
f(n) \propto n^{-\gamma}~~~~T = T_c\:.
\label{fn0}
\end{equation}
We argue that $\xi_{\parallel}$ in equations (\ref{fn-}) and (\ref{fn+}) is
the surface correlation length, below and above the critical
point, respectively. Indeed,
substituting (\ref{fn-}) or (\ref{fn+}) into (\ref{corr1}) and
evaluating the integral by the saddle-point method, we get
\begin{equation}
G_s(l) \propto \exp(-l/\xi_{\parallel})\;,
\label{corr2}
\end{equation}
in accordance with the definition of the surface correlation length.

At the critical point, where $f(n)$ as in (\ref{fn0}), the surface correlation
function in equation (\ref{corr1}) leads to a power law decay
$G_s(l) \sim l^{-4 \gamma}$. Thus the decay exponent, $\eta_{\parallel}$, of 
the critical surface correlation function is given by
\begin{equation}
\eta_{\parallel}=4 \gamma\;.
\label{eta}
\end{equation}
We conclude at this point, that we have obtained a complete
description about the surface properties of the random
Ising model by the star-triangle
method. In the following, we shall use the above formalism to calculate
numerically the surface magnetization, the critical
surface correlations, and the correlation length.

\subsection{Numerical results}

In the actual calculations, we considered the random Ising model 
of the MC simulations, with a (1,1) surface, on a
strip of width $L$ of a diagonal square lattice (which can be considered
as a triangular lattice with vanishing vertical couplings) and imposed
periodic boundary conditions. To reduce finite size effects, we considered
relatively large strips (with $L$ up to 512). We checked that the difference
between the results for the two largest strips ($L=256$ and $L=512$) is 
essentially negligible, doing up to $n=2000$ iterations\cite{note}.
We calculated the
partition function $f(n)$ as a function of $n$, averaging
over several (typically around twenty) random configurations of the couplings.
The ratio $r$ between the two, weak and strong, random couplings was chosen
to be 1, 1/4, and 1/10, as in the simulations; both couplings occur with
the same probability, $p =1/2$.

We start with the analysis of the results in the ordered phase, i.e. $T<T_c$.
For a given temperature, $f(n)$ approaches the surface magnetization
$m_1$, see (\ref{ms}), which is found to agree (within the error of the
calculations) with
MC data presented in the previous Section. Approaching the critical
point, the convergency of $f(n)$ with $n$ becomes slower, in
accordance with the form of the correction term in (\ref{fn-}). Accordingly,
accurate estimates become more difficult. As in the case of the simulations,
the ensemble sampling over different configurations seems to be, however, the
main source of error.

From the values for $m_1(t)$ at different reduced temperatures $t$, we
determined effective surface magnetization exponents
$\beta(1)_{eff}(t)$, as defined in (\ref{betaeff}). The estimates of
the effective exponents obtained from the star-triangle method are
close to those found by the MC technique, see Fig.3. Thus we
confirm that $\beta_1$ is rather robust against introducing randomness
in the two--dimensional Ising model.

At $T_c$, we studied the surface correlation function, as follows from
the partition function $f(n)$. As 
shown in Fig. 6, $f(n)$ exhibits, with $n$ ranging from 100 to 1000, on a
log-log plot ($\ln(f(n))$ vs. $\ln (n)$),
an almost linear behaviour. The average 
slope then defines an average decay
exponent $\gamma_{av}$, see (\ref{fn0}). For the perfect model, our estimate
agrees nicely with the exact value $\gamma_{pure}=1/4$. In the random
case, the average exponent decreases with rising
randomness, i.e. decreasing value of $r$.
For $100 < n <1000$, we obtain the average exponents $\gamma_{av}=0.228$ 
and $\gamma_{av}=0.207$, at $r=1/4$ and $r=1/10$, respectively. Based
on these estimates, one may argue, that also the decay exponent
$\eta_{\parallel}$, see equation (\ref{eta}), varies
with dilution, $r$. However, a more detailed analysis is
needed to substantiate or repudiate these statements. For instance, looking 
at the local effective exponent, defined in analogy to (\ref{betaeff}),
a slight increase of that exponent with
increasing $n$ is observed. Indeed, the
data for $f(n)$ depicted in Fig. 6, may be fitted by the power law of the 
perfect model modified by logarithmic corrections with almost identical
confidence (doing least square fits) as by the power laws with the average,
dilution dependent exponents. Thence, our data
leave room to different interpretations.

In the disordered phase of the model, $T>T_c$, we studied
the correlation length from the asymptotic decay of $f(n)$ in (\ref{fn+}).
Examples of the results of our calculations are shown in Fig. 7, plotting
$\ln(f(n))$ as a function of $n$ at several
temperatures $t$ for $r=1/10$. As seen from that figure, $f(n)$ seems
to exhibit an exponential decay, with $\xi_{\parallel}(t)^{-2}$ following from
the slopes of the curves.
Approaching the critical temperature $T_c$, the correlation length is
expected to diverge as
$\xi_{\parallel}(t) \sim t^{-\nu}$.
From data at $t > 0.05$, we calculated average critical
exponents $\nu_{av}(t)$, with
$\nu_{av}=1.07(2)$ at $r= 1/4$, and $\nu_{av}=1.13(6)$
at $r=1/10$.
These average exponents are larger than the asymptotic exponent
of the pure model, $\nu_{pure}=1$, see
(\ref{xipar}), and vary with the degree of dilution. Note that
similar values have been obtained before by using finite size
scaling \cite{quei2}. However, those average values have been
interpreted as reflecting logarithmic corrections to the
power law of the perfect case \cite{quei2}. Again, we cannot rule out
that possibility.

For further interpretation of
our data, we consider the scaling relation \cite{bind1}
\begin{equation}
\beta_1= \nu \eta_{\parallel}/2\;,
\label{scale}
\end{equation}
which is satisfied, within the errors of the estimates,
by the average exponents, both for $r=1/4$ and $r=1/10$. Following the
alternate interpretation involving logarithmic corrections,  
the critical surface correlations, described by
$\eta_{\parallel}$, would be then affected by
logarithmic corrections, due to the correction terms in the  
correlation length (and their presumed absence in the 
surface magnetization). 
The above scaling law, (\ref{scale}), can be obtained by relating the
surface correlation function between two spins at a distance
of the correlation length, $\xi(t)$, to the square of the surface
magnetization, 
\begin{equation}
G_s(\xi(t)) \sim m_1^2(t)\;,
\label{scale1}
\end{equation}
in the limit $t \to 0$ (when logarithmic corrections are
present, such a relation has been, for instance, checked for
the $q=4$ state Potts model\cite{cardy}). Then, supposing logarithmic
terms in the surface correlations, but not in the surface
magnetization, one easily arrives at the conjecture
\begin{equation}
G_s(l)\sim l^{-1}(\ln l)^{1/2}\;.
\label{conj}
\end{equation}

\noindent
\section{Summary}
In this paper, the boundary critical properties of the two-dimensional
random Ising model have been studied by MC techniques and by the
star--triangle (ST) approach.
In the simulations, we computed magnetization profiles, allowing to
monitor surface and bulk quantities simultaneously. On the other
hand, by the ST method we calculated the surface magnetization as well as
surface correlation functions. Both methods provide data for the
surface magnetization which are in very good agreement, demonstrating
the correctness and accuracy of the two approaches.

To analyse the behaviour of the random Ising model in the critical region,
we considered three singular quantities: the surface magnetization, the
(surface) correlation length and the critical surface correlation function.
The surface magnetization of the dilute model, as obtained from both
methods, follows closely the power law of the corresponding 
perfect model, where $\beta_1$ = 1/2, showing the robustness
of that exponent against even fairly strong randomness.

The behaviour of the other two singular quantities, the critical
surface correlation function and the correlation length, as determined
from the ST method, is rather subtle.
Note that in the ST method we averaged over the {\it logarithm}
of the surface correlation function, leading to information about the
{\it typical} behaviour of the correlation length. The
numerical estimates for the critical exponents of both quantities, i.e.
$\nu_{\parallel}$ and $\eta_{\parallel}$, are found to be dilution dependent,
when calculating average exponents. Similar findings have
been reported before for bulk exponents in the two--dimensional
random Ising model.
These non--universal average bulk exponents have been
interpreted previously either
as reflecting the true asymptotics (satisfying
weak universality) or as being due to logarithmic corrections
of the power laws in the perfect model (in accordance with field-theoretical
predictions). Our numerical data for the surface quantities leave
room to both types of interpretation, as concerns bulk and
surface critical properties. Extending the
field-theoretical predictions and attributing the apparent variation of
the average exponents with the degree of dilution to logarithmic
corrections, we conjectured, in equation (\ref{conj}), the
asymptotic form of the critical surface correlation function.

\section*{Appendix}

We consider the first two layers of a hexagonal Ising lattice (Fig. 4), where
a surface spin $s$ is connected to the second layer spins $s_+$ and
$s_-$ by couplings $p_+$ and $p_-$, respectively. We are interested in a
relation between the thermal averages $\langle s_+ \rangle$,
$\langle s_- \rangle$ and $\langle s \rangle$.

We start by considering the conditional probability
\begin{equation}
\left.P(s)\right|_{s_+,s_-}={\exp(p_+ s s_+ + p_- s s_-)
\over \sum_s \exp(p_+ s s_+ + p_- s s_-)}\;,
\label{prob}
\end{equation}
with fixed values of $s_+$ and $s_-$. Under this condition the
expectational value of $s$ is given by
\begin{eqnarray}
\fl \left.\langle s \rangle \right|_{s_+,s_-}=\tanh(p_+ s s_+ + p_- s s_-)
\nonumber\\
\lo=\tanh\left[(s_+ + s_-){p_+ + p_- \over 2} + (s_+ - s_-){p_+ - p_-
\over 2} \right]\;,
\label{s1}
\end{eqnarray}
which can be evaluated using the fact that $s_+=\pm 1$ and
$s_-=\pm 1$ as
\begin{eqnarray}
\fl \left.\langle s \rangle \right|_{s_+,s_-}=s_+{\tanh(p_+ + p_-) +
\tanh(p_+ - p_-) \over 2}\nonumber\\
\lo+ s_-{\tanh(p_+ + p_-) +
\tanh(p_- - p_+) \over 2}\;.
\label{s2}
\end{eqnarray}
At this point, one can sum over the variables $s_+$ and $s_-$ leading to
\begin{eqnarray}
\fl \langle s \rangle =\langle s_+ \rangle {\tanh(p_+ + p_-) +
\tanh(p_+ - p_-) \over 2}\nonumber\\
\lo+ \langle s_- \rangle {\tanh(p_+ + p_-) +
\tanh(p_- - p_+) \over 2}\;,
\label{s3}
\end{eqnarray}
which is equivalent to equation ({\ref{magnrel}).

The connected surface correlation function of the $n$-th triangular model
$g^{(n)}(i+l,i)=\langle s^{(n)}_{i+l} s^{(n)}_i \rangle - 
\langle s^{(n)}_{i+l} \rangle
\langle s^{(n)}_i \rangle$ and that of the $(n+1)$-th model
are related by
\begin{eqnarray}
\fl g^{(n)}(i+l,i) = a^{(n+1)}_{(i+l)+} a^{(n+1)}_{i+} g^{(n+1)}(i+l+1,i+1)+
a^{(n+1)}_{(i+l)+} a^{(n+1)}_{i-} g^{(n+1)}(i+l+1,i-1)\nonumber\\
\lo+ a^{(n+1)}_{(i+l)-} a^{(n+1)}_{i+} g^{(n+1)}(i+l-1,i+1)\nonumber\\
\lo+ a^{(n+1)}_{(i+l)-} a^{(n+1)}_{i-} g^{(n+1)}(i+l-1,i-1)\;,
\label{g1}
\end{eqnarray}
which can be obtained along the lines of (\ref{magnrel}). Iterating the
expression in (\ref{g1}), one obtains a sum, each term of which can be
formally represented by two directed walks, which start at positions
$i+l$ and $i$, respectively.
If the two walks meet at step $n$ and at some position $j$, then
$g^{(n)}(j,j)=1-\langle s^{(n)}_j \rangle^2$
and the walks annihilate each other. In the $n \to \infty$ limit, the
non-vanishing contribution to $g^{(0)}(i+l,i)=g(i+l,i)$ is given by those
processes, which are connected to annihilated walks. In the
transfer matrix notation the average surface correlation function is
given by
\begin{equation}
G_s(l)={1 \over L} \sum_{i=1}^L \sum_{a.w.} \left[ \prod_{m=1}^n {\bf A}^{(m)}
\right]_{i+l,j}\left[ \prod_{k=1}^n {\bf A}^{(k)}
\right]_{i,j} \left[1-\left(\sigma_j^{(n)}\right)^2\right]\;.
\label{g2}
\end{equation}

The asymptotic behavior of this expression can be obtained by noticing that
the transverse fluctuations of directed walks have Gaussian nature, and the
corresponding probability distribution is sharp. Consequently, in the large
$l$ (and large $n$) limit it is enough to consider the typical processes. Then
there are two factors in (\ref{g2}), which are both
approaching the partition function
of $n$-step directed walks, $f(n)$, see (\ref{fn}), and these contributions
should be multiplied by $P_n(l)$, the ratio of those walks which
are annihilated
at the $n$-th step. In this way, we obtain
\begin{equation}
G_s(l) \approx \sum_n P_n(l)[f^2(n)-f^2(\infty)]\;,
\label{g3}
\end{equation}
which in the continuum approximation is given in (\ref{corr1}).

\ack
P.L. and F. Sz. would like to thank the Institutes for Theoretical Physics
at the Universit\"at Hannover, especially Prof. H. U. Everts, and
at the Technische Hochschule Aachen for
kind hospitality, as well as 
the Deutsche Akademische Austauschdienst and the Soros
Foundation, Budapest, for facilitating their visits. F.I.'s work has been
supported by the Hungarian National Research Fund under grants OTKA TO12830
and OTKA TO23642 and by the Ministery of Education under grant
No FKFP 0765/1997.

\Figures
\begin{figure}
\caption{Magnetization profiles $m(i)$ of two--dimensional
perfect (squares) and random, $r=1/4$ (circles), Ising 
models with (11) surfaces, at $t=0.2$ (open symbols) and $t=0.05$
(full symbols). Systems of size 160$\times$80 were simulated.}
%\label{fig1}
\end{figure}

\begin{figure}
\caption{Effective exponent $\beta(i)_{eff}$, with i=1,2,3, and 10,
vs. reduced temperature $t$, for
the perfect Ising model with (11) surface. The solid lines 
denote exact results[14,22]. Monte Carlo data for systems of sizes 
80$\times$40 ($t > 0.3$), 160$\times$80 ($0.07 < t < 0.3$), and
320$\times$160 ($t < 0.07$) are shown.}
%\label{fig2}
\end{figure}

\begin{figure}
\caption{Effective exponent $\beta(1)_{eff}$, vs. reduced temperature
$t$ for the random two--dimensional Ising model with (11) surface, at
$r$=1/4 (circles) and $r$=1/10 (triangles). Systems of sizes
80$\times$40 ($t > 0.3$), 160$\times 80$ ($t=0.275$), 320$\times$160
($0.1 <t < 0.275$), 640$\times$320 ($0.05 < t <0.1$), and
1280$\times$640 ($t < 0.05$) were simulated. The solid line denotes
the exact result in the perfect case[22].}  
%\label{fig3}
\end{figure}

\begin{figure}
\caption{Mapping of the original triangular lattice (solid line) to an
equivalent hexagonal lattice (dashed line) and further to a new triangular
lattice (dotted line) using the star-triangle transformation. The surface
spins of the $n$-th model $(S_l^{(n)})$ and those of the $(n+1)$-th
model $(S_{l+}^{(n+1)},~S_{l-}^{(n+1)})$ are connected by the surface
couplings of the intermediate $(n+1)$-th hexagonal model $(p_{l+}^{(n+1)},
~p_{l-}^{(n+1)})$. The couplings $\{K_i\}$, $\{p_i\},~i=1,2,3$ appearing
in the star-triangle relation in (\protect\ref{transf}) are also indicated.}
%\label{fig4}
\end{figure}

\newpage
\begin{figure}
\caption{Magnetization of the $(1,1)$ surface of the two-dimensional random
Ising model with $r=1/10$ as a function of the temperature. The finite
iteration approximants $f(n)$ of the star-triangle method in
equation (\protect\ref{fn})
are indicated by circles $(n=128)$, squares $(n=256)$, triangles $(n=512)$ and
by crosses $(n=1024)$. The asymptotic behaviour of $f(n)$ is different for
$T<T_c$, $T>T_c$ and at $T=T_c$, as given in (\protect\ref{fn-}),(\protect\ref{fn+}) and
(\protect\ref{fn0}), respectively.}
%\label{fig5}
\end{figure}

\begin{figure}
\caption{Finite iteration approximants to the surface magnetization, $f(n)$,
as a function of $n$ in a log-log plot, at the critical point of the
two-dimensional random Ising model with dilution $r=1/10$ (squares) and
$r=1/4$ (circles), compared with the perfect model (triangles). The
slope of the curves, indicated by straight lines, is related to the
average decay exponent $\eta_{\parallel}$ of the critical
surface correlations
through (\protect\ref{eta}), see text.}
%\label{fig6}
\end{figure}

\begin{figure}
\caption{Finite iteration approximants to the surface magnetization, $f(n)$, 
as a function of $n$ in a semi-logarithmic plot, at different reduced
temperatures $t=0.1$ (triangles), $t=0.2$ (circles) and $t=0.3$ (squares) above
the critical point of the two-dimensional random Ising model with
$r=1/10$. The slope of the curves, indicated by
straight lines, corresponds to the
inverse square of the average correlation length, see
equation (\protect\ref{fn+}).}
%\label{fig7}
\end{figure}


\begin{thebibliography}{99}
\bibitem{note} One expects that the characteristic values of $n$ and $L$
are related by $L \sim n^{1/2}$, like the size of transverse
fluctuations of a directed walk to the number $n$ of steps
\end{thebibliography}
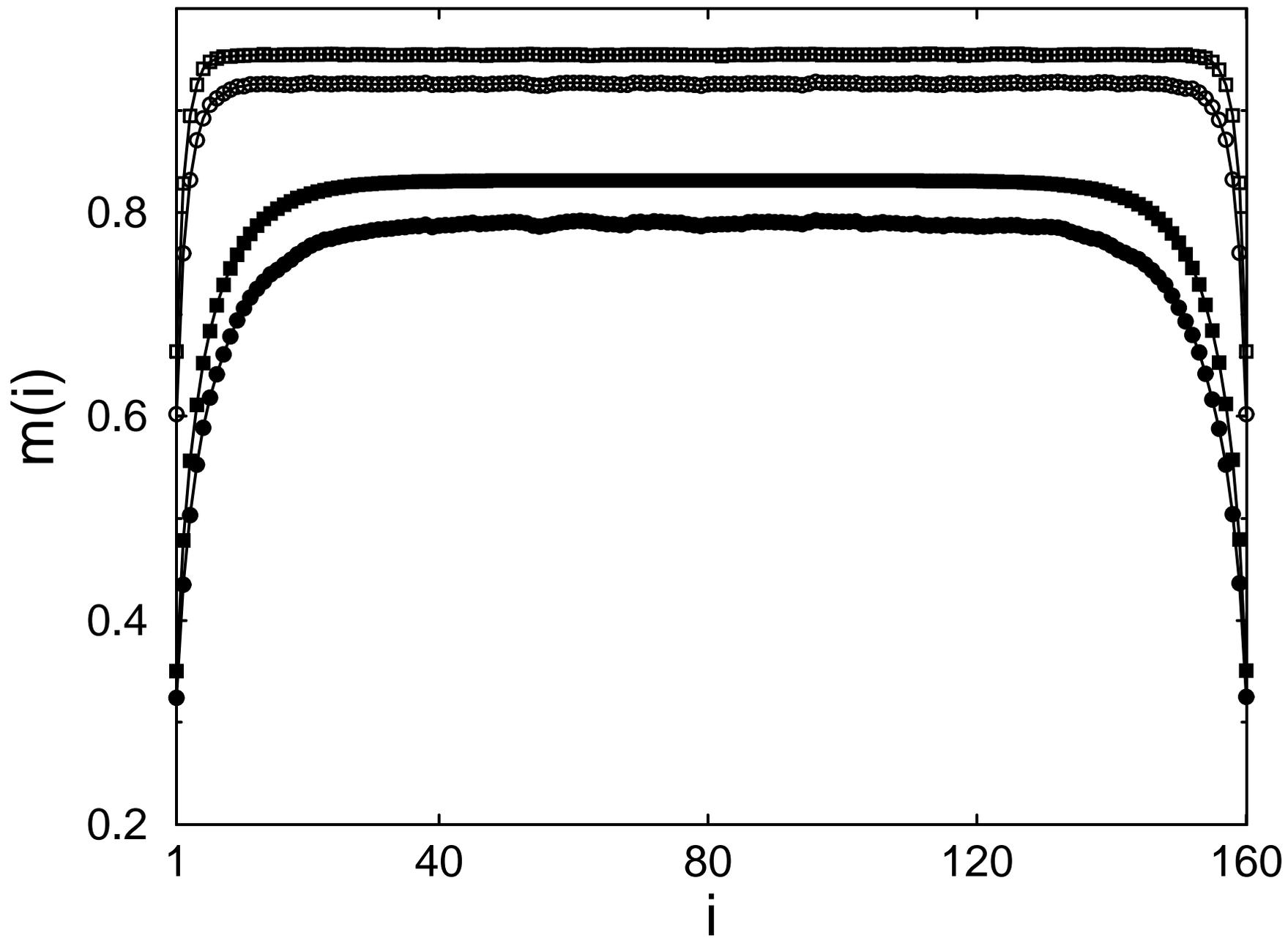
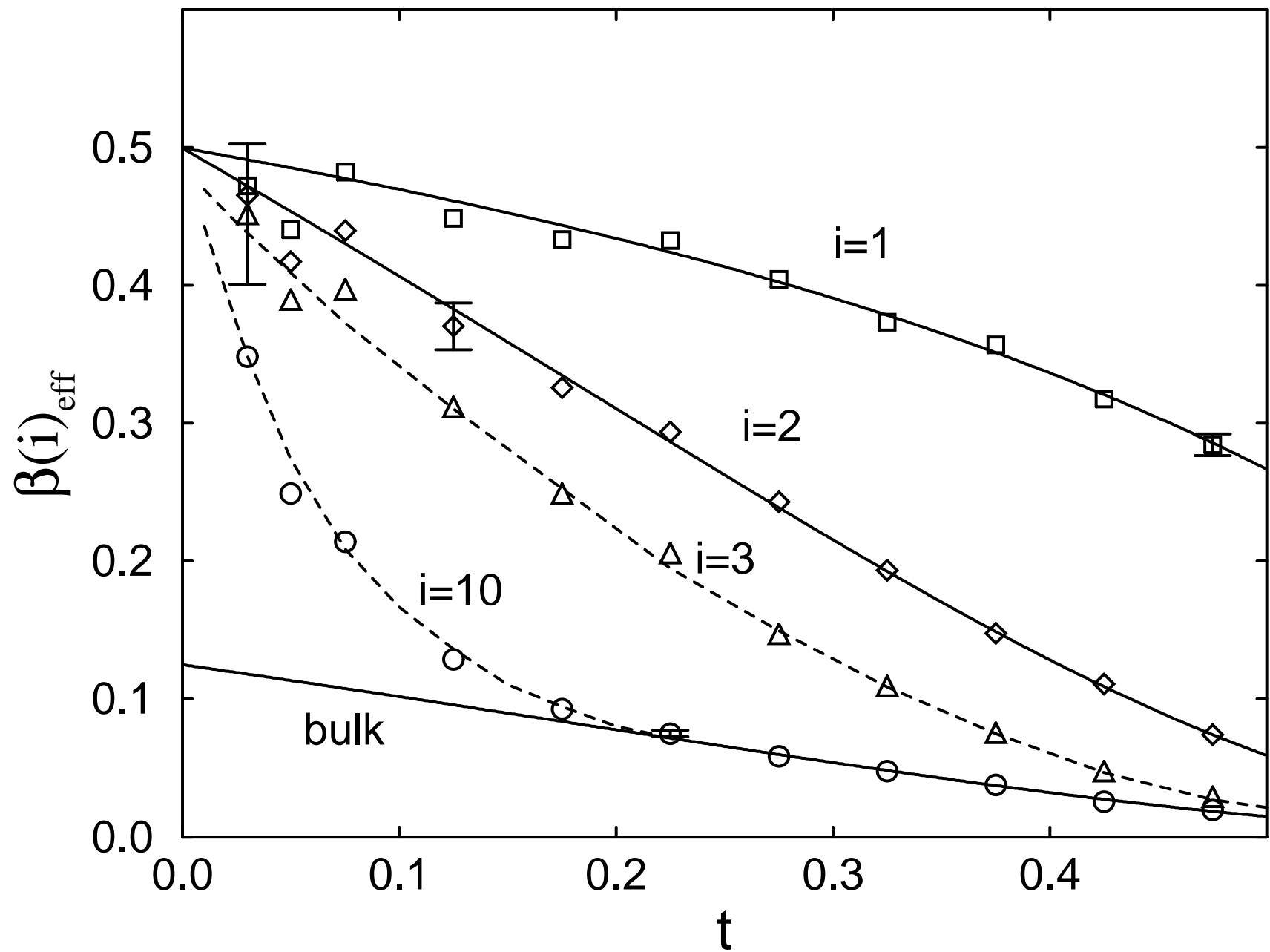
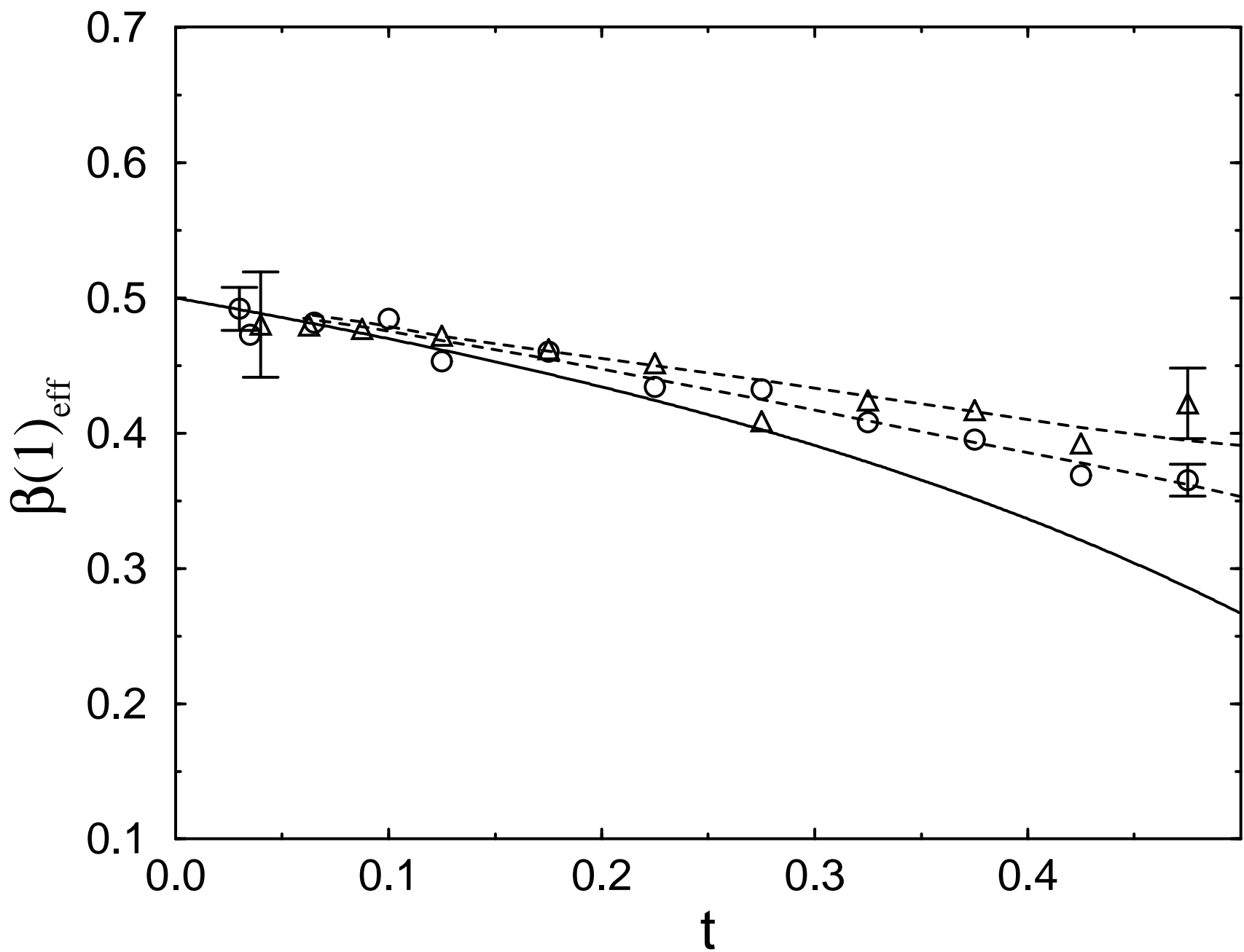
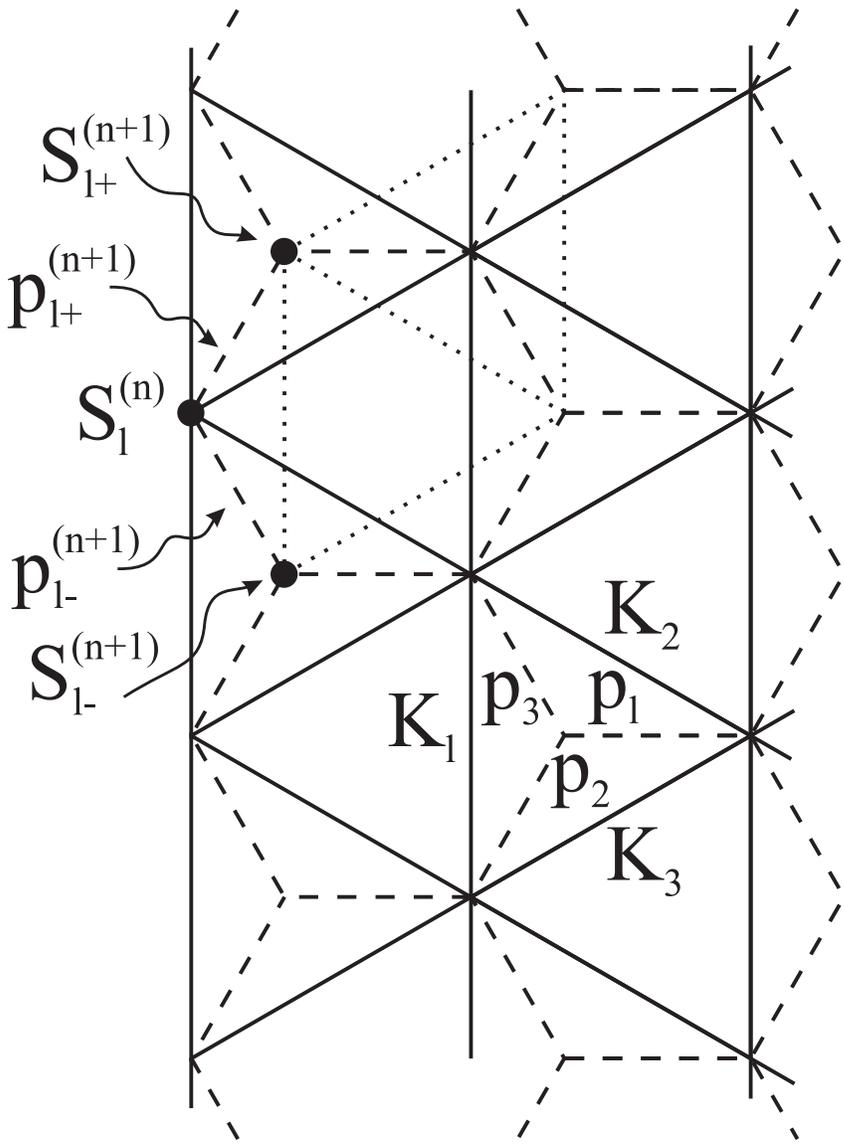
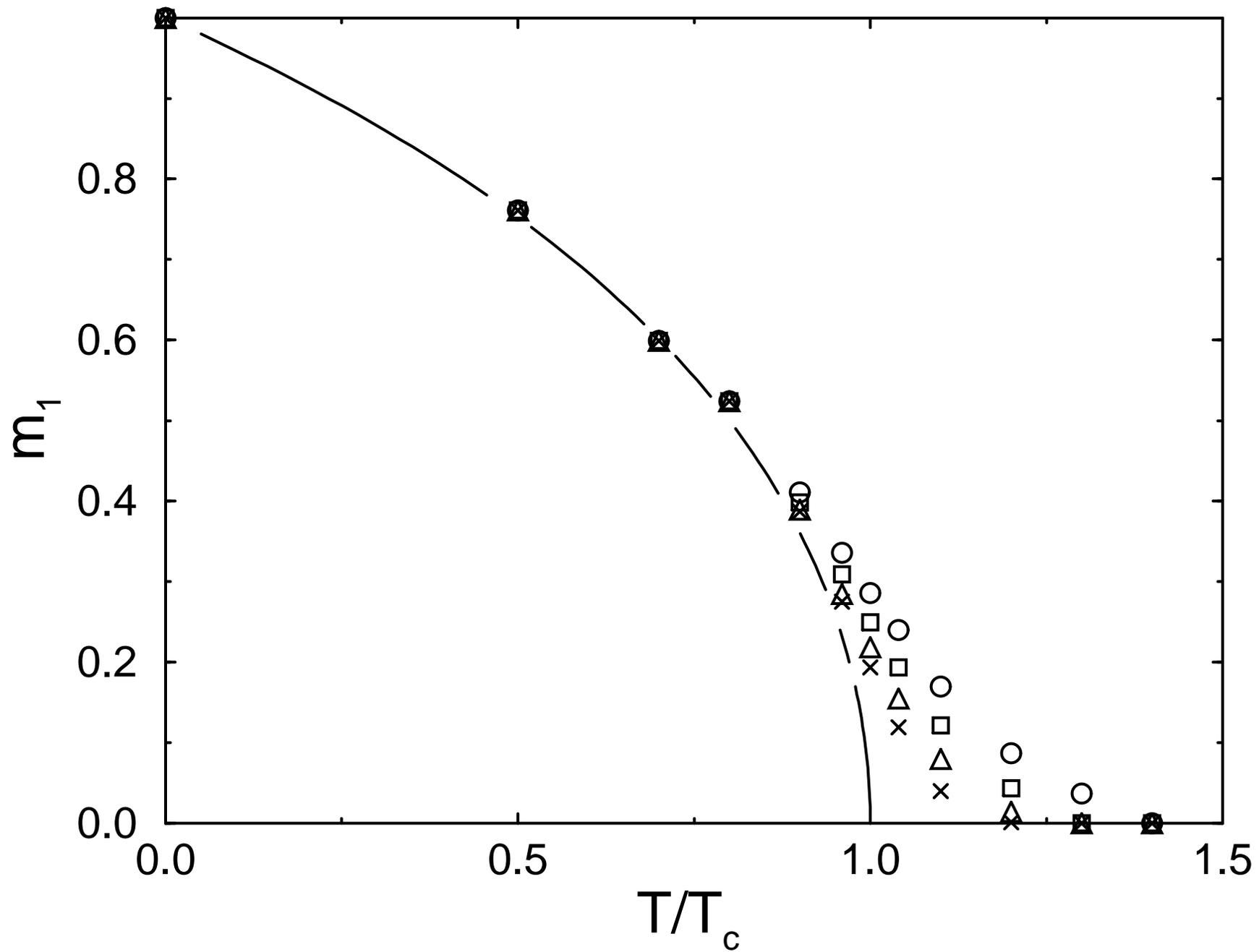
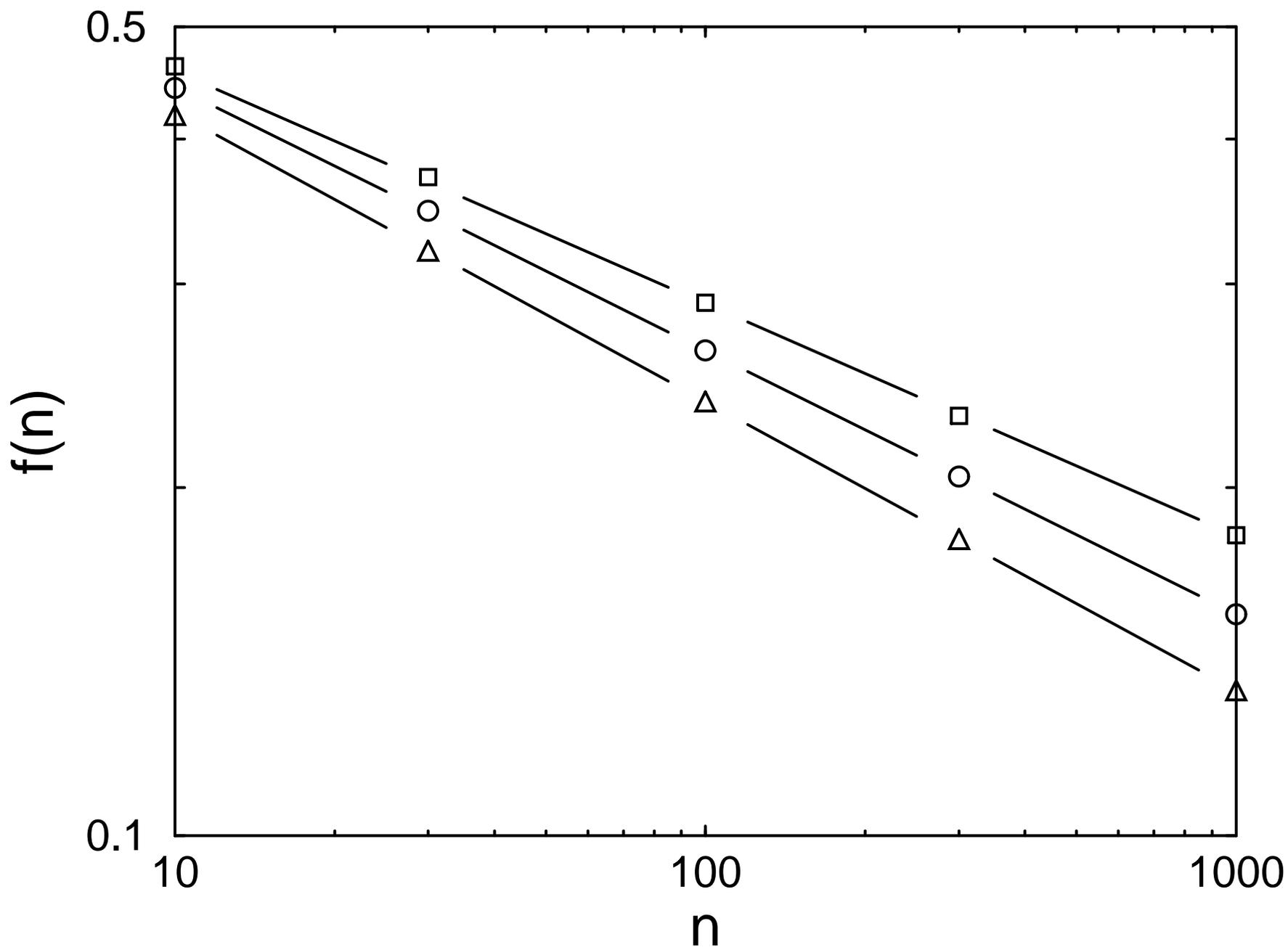
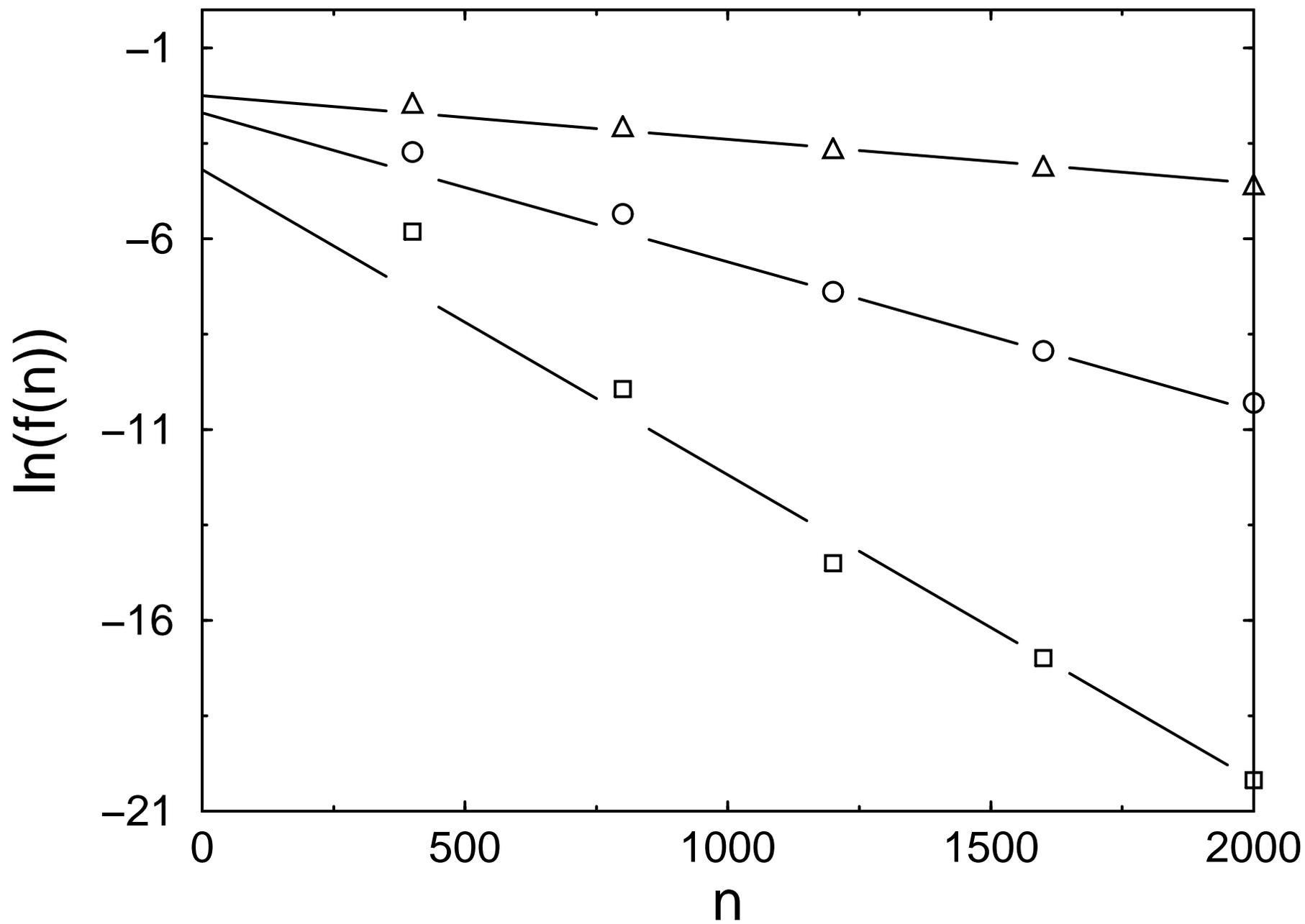
\end{document}